\documentclass[preprint]{revtex4-1}
\usepackage{graphicx}

\def\be{\begin{equation}}	\def\ee#1{\label{#1}\end{equation}}
\def\ba{\begin{array}}	\def\ea{\end{array}}
\def\bea{\begin{eqnarray}}	\def\eea{\end{eqnarray}}
\def\mc{\mathcal}		\def\pa{\partial}
  \def\ci{\cite}  
\def\ra{\rightarrow}		
	
\def\N{{\mbox{\scriptsize N}}}	\def\S{{\mbox{\scriptsize S}}}
			\def\L{{\mbox{\scriptsize L}}}
\def\TOT{{\mbox{\scriptsize TOT}}} \def\S{{\mbox{\scriptsize S}}}
\def\QCD{{\mbox{\scriptsize QCD}}} 
 
\def\Re{{\rm Re}}   \def\Im{{\rm Im}}   

\def\Title#1{\begin{center} {\Large #1 } \end{center}}
\def\Author#1{\begin{center}{ \sc #1} \end{center}}
\def\Address#1{\begin{center}{ \it #1} \end{center}}

\newcommand\pubnumber{MAXLA-3/19, ``Salpeter''}
\newcommand\pubdate{\today}
\newcommand\pubblock{\rightline{\begin{tabular}{l} \pubnumber\\
\pubdate \end{tabular}}}

\newenvironment{Presented}
{\begin{quotation} 
\begin{center} 
 PRESENTED AT\end{center}  
\begin{center}\begin{large}}{\end{large}
\end{center} 
\end{quotation}}

\begin{document}

\begin{titlepage}
\pubblock

\vfill
\Title{ Transformation of the spinless Salpeter equation} 
\vspace{5mm}
\Author{Mikhail N. Sergeenko}
\Address{Fr. Skaryna Gomel State University, 
BY-246019, Gomel, BELARUS\\
{\rm msergeen@usa.com}}
\vfill

\begin{abstract}
 Spinless Salpeter equation for two bound particles is analized. 
 We use the fact that in relativistic kinematics the spatial two 
particle relative momentum is relativistic invariant. 
 Free particle hypothesis for the bound state is developed: 
comstituents move as free particles inside of the system. 
 The Shr\"odinger-type wave equation is derived. 
 Three equivalent forms of the eigenvalue equation are given. 
 Relative motion of quarks in eigen states is described by the 
asymptotic solution in the form of the standing wave of $\cos(kx+a)$ 
for each spatial degree of freedom. 
 To test the model the spin center-of-gravity energy levels for 
the hydrogen atom are calculated and compared with the NIST data. 
 Complex eigenmasses for the $H$ atom are obtained. 

%
\vskip 5mm
\noindent Pacs: 11.10.St; 12.39.Pn; 12.40.Nn; 12.40.Yx
\end{abstract}

\vspace{3mm}
\begin{Presented}
NONLINEAR PHENOMENA IN COMPLEX SYSTEMS \\
XXVI  International  Seminar\\
Chaos, Fractals, Phase Transitions, Self-organization \\
May 21--24, 2019, Minsk, Belarus 
\end{Presented}
\vspace{5mm}
\end{titlepage}

\section{Introduction}\label{intro}
 Relativistic few-body problem has received a great attention in many fields of researches including particle physics. 
 Solution of the problem in a~way fully consistent with all requirements 
imposed by special relativity and within the framework of Quantum Field 
Theory (QFT) is one of the great challenges in theoretical elementary particle physics. 
 The most complete results here exist for the case of two particles. 
 However, even the simplest relativistic two-body (R2B) bound-state 
problem has several principal difficulties.  

 There are two main aspects of the problem: 
1)~the scattering of two particles and 2)~bound state of the particles. 
 These are two main R2B problems in quantum physics. 
 The scattering and bound-state problems in quantum physics are related 
to solution of a~single covariant equation, but for different boundary 
conditions. 
 The R2B bound-state problem has several principal difficulties. 
 Main of them concern $\alpha)$~the equation of motion and 
$\beta)$~the interaction potential. 
 Judging from the large variety of approaches attempted even 
in recent years, this problem has no generally agreed-upon 
solution~\ci{Weinb95,LuchSho16}. 

 Description of the R2B bound systems has a~long history. 
 There have been proposed several wave equations for the describtion 
of bound states within relativistic quantum theory. 
 Solution of the~R2B problem in a~way fully consistent with all 
requirements imposed by special relativity and within the~framework 
of QFT is founded on the four-dimensional (4D) covariant Bethe-Salpeter 
equation (BS)~\ci{BethSal08,SalBeth51}. 
 The better-known work is an~integral BS equation in momentum space 
that is obtained directly from QFT. 

 The 4D covariant BS equation governs all the bound states and 
the scattering in R2B problem; it is appropriate framework 
for the description of the R2B bound-state problem within 
QFT~\ci{LuchSho16}. 
 However, the BS equation cannot be solved in general. 
 Many attempts to apply the BS formalism to the R2B bound-state 
problem give series of difficulties, such as the relative time problem, 
the impossibility to determine the BS interaction kernel beyond 
the tight limits of perturbation theory and many others~\ci{LuchSho99}. 

 There exist various reductions of the BS 
equation~\ci{Salpet52,Kadysh68,Nakan69,Todor76,Sutt79,LuchSho16}. 
 For a~variety of reasons most of the attempts are not appropriate 
for the treatment of highly relativistic effects like those necessary 
for the calculation of bound states. 
 Many authors have developed noncovariant instantaneous truncations 
of the BS equation~\ci{Nakan69,LuchSho16}. 
 A~better known is the Salpeter work~\ci{Salpet52}. 
 These and many other difficulties are the sources of the numerous 
attempts to reformulate the R2B 
problem~\ci{Kadysh68,Nakan69,JallSaz97,Crater00,LuchSho16}. 
 This rather long list of authors and papers include different 
formulations of the R2B problem and related applications to QED and 
QCD. 
 Most of the references are related to constraint 
dynamics~\ci{Crater00}. 

 There are a number of strategies in computational treatments of 
QCD that emerge in the study of meson spectroscopy. 
 One is to set up a discrete lattice analog of the full QFT. 
 Another is to first make analytic approximations which replace 
the QFT problem by a classical variational problem involving 
an effective Lagrange function and action. 
 The latter approach has been exploited and gave a detailed account 
of applications of the R2B Dirac equations of constraint dynamics 
to the meson quark-antiquark bound states~\ci{CratVanA04}. 
 Applications of R2B Dirac Equations to the meson spectrum with 
three versus two covariant interactions, SU(3) mixing, and
comparison to a quasipotential approach were considered 
in~\ci{CratSche10}. 
 The R2B Dirac equations of Constraint Dynamics have dual origins. 
 On the one hand they arise as one of the many quasipotential 
reductions of the BS equation. 
 On the other they arise independently from the development of 
a consistent covariant approach to the R2B problem in relativistic 
classical mechanics independent of QFT~\ci{CratVanA83}. 

 In this work we consider these two aspects and then go on to discuss applications to the Hydrogen atom and hadron spectroscopy. 
 We start with R2B problem in relativistic classical mechanics
 Using relativistic kinematics and the correspondence principle, 
we deduce a two-particle wave equation. 
 The interaction of particles (quarks) is described by the modified 
funnel-type Lorentz-scalar Cornell potential. 
 We obtain two exact asymptotic solutions of the equation which 
are used to write the complex-mass formula for the bound system. 
 The last part of the work explains the importance we put on 
numerical tests of the model and some speculative theoretical 
results concerning the Hydrogen atom.

\section{ Quasipotential Reduction of the BS Equation}\label{Rel2B}
 The homogeneous BS equation governs all the bound states. 
 However, numerious attempts to apply the BS formalism to relativistic 
bound-state problems give series of difficulties. 
 Its inherent complexity usually prevents to find the exact solutions 
or results in the appearance of excitations in the relative time 
variable of the bound-state constituents (abnormal solutions), 
which are difficult to interpret in the framework of quantum 
physics~\ci{LuchSho99}. 
 Usually, calculations are carried out with the help of 
phenomenological and relativistic models~\ci{Morp90,EbFausGa11}. 

 The BS equation~\ci{BethSal08,SalBeth51}, which is the basic bound state equation in QFT, has been revealed inadequate for quantitative calculation. 
 In practice, the BS equation has been used in QED in the Coulomb 
gauge, which is a noncovariant gauge. 
 Because of the instantaneous nature of the dominant part of the photon 
propagator, one is able to transform the original 4D  equation into 
a 3D one and to avoid the previous difficulties~\ci{BodYen78,BuchRem80}. 
 However, the latter gauge has its own limitations. 
 It neccesitates a different treatment of exchanged photons and of 
photons entering in radiative corrections. 
 Additional complications arise when QED is mixed with other 
interactions, where already covariant propagators are present.
 In this respect, the wave equations obtained in the framework of constraint theory~\ci{LongLus86,JallSaz97,Todor76,Crater00} have been 
shown to provide a satisfactory answer to the requirement of 
a covariant treatment of perturbation theory in the bound state 
problem~\ci{JallSaz97}. 

 More valuable are methods which provide either exact or approximate 
analytic solutions for various forms of differential equations. 
 They may be remedied in three-dimensional reductions of the BS 
equation. 
 In most cases the analytic solution can be found if original 
equation is reduced to the Schr\"odinger-type wave equation. 
 The most well-known of the resulting bound-state equations is 
the one proposed by Salpeter~\ci{Salpet52}. 
 There exist many other approaches to bound-state problem. 

 Two body BS equation~\ci{BethSal08,SalBeth51,LuchSho16} for spin-zero bound states is
\be
G_0^{-1}\Psi \equiv (p_1^2 + m_1^2)(p_2^2 + m_2^2)\Psi = K\Psi,
\ee{BSzer}
where $G_0=G_{0,1}G_{0,2}$ is free propagator of particles. 
 The irreducible BS kernel $K$ would in general contain charge renormalization, vacuum polarization graphs and could contain s
elf-energy terms transferred from the inverse propogators.
 The kernel $K$ is obtained from the off-mass-shell scattering 
amplitude, 
\be
T = K + KG_0T.
\ee{ScattAmp}
 Recent work with static models has indicated, that abnormal solutions 
disappear if one includes all ladder and cross ladder 
diagrams~\ci{JallSaz97}. 
 This supports Wick\`\,s conjecture on defects of ladder approximations. 
 In the mean time numerous 3D quasipotential reductions of 
the BS equation had been proposed. 

 Reductions of the BS equation can be obtained from iterating this 
equation around a 3D Lorentz invariant hypersurface in relative 
momentum ($p$) space. 
 This leads to invariant 3D wave equations for relative motion. 
 The resultant 3D wave equation is not unique, but depends on 
the nature of the 3D hypersurface. 
 One can choose Todorov\`\,s quasipotential equation~\ci{Todor76} 
which has this Schr\"odinger-like form
\be
[p^2+\Phi(x_1-x_2)]\psi = \kappa^2(w)\psi,
\ee{TodorEq}
where the quasipotential $\Phi$ is related to the scattering amplitude, 3D hyperfine restriction on the relative momentum $p$ is defined by 
$p\cdot P\psi=0$, $P=p_1+p_2$. 
 The effective eigenvalue in (\ref{TodorEq}) is 
\be
\kappa^2(w) = \frac 1{4w^2}[w^2-(m_1-m_2)^2][w^2-(m_1+m_2)]^2,
\ee{EffMom2}
with $w=\sqrt{P^2}$ the c.m. invariant energy.  

 If one uses a scheme that adapts Eikonal approximation for ladder, cross ladder, and constraint diagrams to bound states applied through 
all orders, it gives for scalar exchange the quasipotential
\be
\Phi = 2m_w S + S^2,
\ee{ScalPot}
while for vector exchang
\be
\Phi = 2\epsilon_w A - A^2. 
\ee{VectPot}
 The kinematical variables
\be
m_w = \frac{m_1 m_2}w,
\ee{RelMass}
\be
\epsilon_w = \frac{w^2 - m_1^2 - m_2^2}{2w},
\ee{RelEner}
satisfy the Einstein relation
\be
\kappa^2(w) = \epsilon_w^2 - m_w^2,
\ee{EinRel}
and corresponds to the energy and reduced mass for the fictitious particle of relative motion. 
 The effects of ladder and cross ladder diagrams thus embedded in 
their c.m. energy dependencies. 

 These forces $\Phi$ to depend on $x_1 - x_2$ only through 
the transverse component, $x_\perp^\mu$. 
 Thus, in the c.m. frame, the hypersurface restriction $p\cdot P\psi=0$  not only eliminates the relative energy [$p\psi=(0,\mathbf p)\psi=0$] 
but implies that the relative time does not appear 
[$x_\perp^\mu=(0,\mathbf r)$].

\section{ The interaction potential}\label{IntPot}
 The nonrelativistiv (NR) quantum mechanics shows very good results 
in describing bound states; this is partly bacause the potential is 
NR concept. 
 In relativistic mechanics one faces with different kind of 
speculations around the potential, because of absence of a strict 
definition of the potential in this theory. 
 In NR formulation, the $H$ atom, for example, is described by 
the Schr\"odinger equation and is usually considered as an electron 
moving in the external field generated by the proton static electric 
field given by the Coulomb potential. 
 In relativistic case, the binding energy of an electron in a~static 
Coulomb field (the external electric field of a point nucleus of 
charge $Ze$ with infinite mass) is determined predominantly by 
the Dirac eigenvalue~\ci{MohrTayl}. 
 The spectroscopic data are usually analyzed with the use of 
the Sommerfeld's fine-structure formula~\ci{Bohm79}, 

 One should note that, in these calculations the $S$ states start 
to be destroyed above $Z=137$, and that the $P$ states being 
destroyed above $Z=274$. 
 Similar situation we observe from the result of the Klein-Gordon 
wave equation, which predicts $S$ states being destroyed above $Z=68$ 
and $P$ states destroyed above $Z=82$. 
 Besides, the radial $S$-wave function $R(r)$ diverges as $r\ra 0$. 
 These problems are general for all Lorentz-vector potentials which 
have been used in these calculations~\ci{Huang01}. 
 In general, there are two different relativistic versions: 
the potential is considered either as the zero component of 
a~four-vector, a~Lorentz-scalar or their mixture~\ci{SahuAll89}; 
its nature is a~serious problem of relativistic potential 
models~\ci{Sucher95}. 

 This problem is very important in hadron physics where, for 
the vector-like confining potential, there are no normalizable 
solutions~\ci{Sucher95,SemayCeu93}. 
 There are normalizable solutions for scalar-like potentials, but not 
for vector-like. 
 This issue was investigated in~\ci{MyZPhC94,Huang01};  
it was shown that the effective interaction has to be Lorentz-scalar 
in order to confine quarks and gluons. 
 The relativistic correction for the case of the Lorentz-vector 
potential is different from that for the case of the Lorentz-scalar 
potential~\ci{MyMPLA97}. 

 Quarkonia as quark-antiquark bound states are simplest among mesons. 
 The quarkonium universal mass formula and ``saturating'' Regge 
trajectories were derived in~\ci{MyZPhC94} and in~\ci{MyEPJC12,MyEPL10} applied for gluonia (glueballs). 
 The mass formula was obtained by interpolating between NR heavy 
$Q\bar Q$ quark system and ultra-relativistic limiting case of light 
$q\bar q$ mesons for the Cornell potential~\ci{LattBali01,EichGMR08}, 
\be 
V(r)=V_\S(r)+V_\L(r)\equiv-\frac 43\frac{\alpha_\S}r +\sigma r.
\ee{CornPot} 
 The short-range Coulomb-type term $V_\S(r)$, originating from one-gluon exchange, dominates for heavy mesons and the linear one $V_\L(r)$, which 
models the string tension, dominates for light mesons. 
 Parameters $\alpha_\S$ and $\sigma$ are directly related to basic 
physical quantities of mesons. 

 The Cornell potential (\ref{CornPot}) is fixed by the two free 
parameters, $\alpha_\S$ and $\sigma$. 
 However, the strong coupling $\alpha_\S$ in QCD is a~function 
$\alpha_\S(Q^2)$ of virtuality $Q^2$ or $\alpha_\S(r)$ in configuration 
space. 
 The potential can be modified by introducing the 
$\alpha_\S(r)$-dependence, which is unknown. 
 A~possible modification of $\alpha_\S(r)$ was introduced in~\ci{MyEPJC12}, 
\be
V_\QCD(r) = -\frac 43\frac{\alpha_\S(r)}r +\sigma r,\quad 
\alpha_\S(r)=\frac 1{b_0\ln[1/(\Lambda r)^2+(2\mu_g/\Lambda)^2]},
\ee{VmodCor}
where $b_0=(33-2n_f)/12\pi$, $n_f$ is number of flavors, 
$\mu_g=\mu(Q^2)$ --- gluon mass at $Q^2=0$, $\Lambda$ is the QCD scale 
parameter. 
the running coupling $\alpha_\S(r)$ in (\ref{VmodCor}) is frozen at  
$r\ra\infty$, $\alpha_\infty=\frac 12[b_0\ln(2\mu_g/\Lambda)]^{-1}$, 
and is in agreement with the asymptotic freedom properties, i.\,e., 
$\alpha_\S(r\ra 0)\ra 0$. 

 In this work we consider and analize general coordinate-space 
relativistic spinless Salpeter (SS) equation for two-body 
system~\ci{Salpet52}. 
 In the c.m. frame, the SS equation has the form ($\hbar=c=1$)
\be
\left[\sqrt{(-i\vec\nabla)^2 + m_1^2} + \sqrt{(-i\vec\nabla)^2 +
m_1^2} + V(r)\right] = E\psi (\vec r) = 0, 
\ee{SSeq}
where $V(r)$ is the potential (for simplicity we consider separable
spherically symmetric potential). 
 It is a problem to find the analytic solution of this equation; 
it can not be reduced to the second-order differential equation of 
the Shr\"odinger type. 
 The problem originates from two square root operators which cause 
a serious difficulties.


\section{ Transformation of the $SS$ equation}\label{TransSS}
 Standard relativistic approaches 
for R2B systems run into serious difficulties in solving known 
relativistic wave equations. 
 Consider the problem in Relativistic Quantum Mechanics (RQM). 
 The formulation of RQM differs from NR QM by the replacement of 
invariance under Galilean transformations with invariance under 
Poincar\`e transformations. 
 The RQM is also known in the literature as relativistic Hamiltonian 
dynamics or Poincar\`e-invariant QM with direct interaction~\ci{Dirac49}. 
 There are three equivalent forms in the RQM called ``instant'', 
``point'', and ``light-front'' forms. 

 The dynamics of many-particle system in the RQM is specified 
by expressing ten generators of the Poincar\`e group, 
$\hat M_{\mu\nu}$ and $\hat W_\mu$, in terms of dynamical variables. 
 In the constructing generators for interacting systems it is customary 
to start with the generators of the corresponding non-interacting 
system; the interaction is added in the way that is consistent with 
Poincare algebra. 
 In the relativistic case it is necessary to add an interaction $V$ 
to more than one generator in order to satisfy the commutation 
relations of the Poincar\'e algebra. 

 The interaction of a~relativistic particle with the 4-momentum 
$p_\mu$ moving in the external field $A_\mu(x)$ is introduced in 
QED according to the gauge invariance principle, 
$p_\mu\ra P_\mu=p_\mu-eA_\mu$. 
 The description in the ``point'' form of RQM implies that the mass 
operators $\hat M^{\mu\nu}$ are the same as for non-interacting 
particles, i.\,e., $\hat M^{\mu\nu}=M^{\mu\nu}$, and these 
interaction terms can be~presented only in the form of 
the 4-momentum operators~$\hat W^\mu$~\ci{MyRQMAnd99}. 

 Consider the R2B problem in classic relativistic theory. 
 Two particles with 4-momenta $p_1^\mu$, $p_2^\mu$ and 
the interaction field $W^\mu(q_1,\,q_2)$ together compose a~closed conservative system, which can be characterised by the 4-vector 
$\mc{P}^\mu$, 
\be 
\mc{P}^\mu = p_1^\mu + p_2^\mu + W^\mu(q_1,\,q_2),
\ee{Main4vec}
where the space-time coordinates $q_1^\mu$, $q_2^\mu$ and 
4-momenta $p_1^\mu$, $p_2^\mu$ are conjugate variables, 
$\mc{P}_\mu \mc{P}^\mu=\mathsf{M}^2$; here $\mathsf{M}$ is 
the system's invariant mass. 
 Underline, that no external field and each particle 
of the system can be considered as moving source of the interaction 
field; the interacting particles and the potential are a~unified 
system. 
 There are the following consequencies of (\ref{Main4vec}) and 
they are key in our approach. 

 The 4-vector (\ref{Main4vec}) describes {\it free motion} of 
the bound system and can be presented as, 
\bea
E = \sqrt{\mathbf{p}_1^2+m_1^2}+\sqrt{\mathbf{p}_2^2+m_2^2}
+W_0(q_1,\,q_2)=\rm{const}, \quad \label{TwoEnr}\\ 
\mathbf{P}=\mathbf{p}_1+\mathbf{p}_2
+\mathbf{W}(q_1,\,q_2)=\rm{const}, \quad \label{TwoMom}
\eea
describing the energy and momentum conservation laws. 
 The energy (\ref{TwoEnr}) and total momentum (\ref{TwoMom}) 
of the system are the constants of motion. 
 By definition, for conservative systems, the intergals (\ref{TwoEnr}) 
and (\ref{TwoMom}) can not depend on time explicitly. 
 This means the interaction $W(q_1,\,q_2)$ should not depend on 
time, i.\,e., $W(q_1,\,q_2)=>V(\mathbf{r}_1,\,\mathbf{r}_2)$. 

 It is well known that the potential as a~function in 3D-space 
is defined by the pro\-pa\-ga\-tor $D(\mathbf{q}^{\,2})$ (Green 
function) of the virtual particle as a carrier of interaction, where 
$\mathbf{q}=\mathbf{p}_1-\mathbf{p}_2$ is the transfered momentum. 
 In case of the Coulomb potential the propagator is 
$D(\mathbf{q}^{\,2})=-1/\mathbf{q}^{\,2}$; the Fourier transform 
of $4\pi\alpha D(\mathbf{q}^{\,2}$) gives the Coulomb potential, 
$V(r)=-\alpha/r$. 
 The relative momentum $\mathbf{q}$ is conjugate to the relative vector 
$\mathbf{r}=\mathbf{r}_1-\mathbf{r}_2$, therefore, one can accept 
that $V(\mathbf{r}_1,\,\mathbf{r}_2)=V(\mathbf{r})$~\ci{LuchSho99}. 
 If the potential is spherically symmetric, one can write  
$V(\mathbf{r})=>V(r)$, where $r=|\mathbf{r}|$.  
 Thus, the system's relative time $\tau=t_1-t_2=0$ 
(instantoneous interaction). 

 Equations (\ref{TwoEnr}) and (\ref{TwoMom}) in the c.m. frame are 
\bea 
\mathsf{M} = 
\sqrt{\mathbf{p}^2+m_1^2}+\sqrt{\mathbf{p}^2 +m_2^2}+\mathsf{V}(r),
\label{ClasE2B} \\
\mathbf{P}=\mathbf{p}_1 +\mathbf{p}_2 +
\mathbf{W}(\mathbf{r}_1,\,\mathbf{r}_2)=\mathbf{0}, \label{ClasM2B}
\eea
where $\mathbf{p}=\mathbf{p}_1 =-\mathbf{p}_2$ that follows from the 
equality $\mathbf{p}_1+\mathbf{p}_2=0$; this means that 
$\mathbf{W}(\mathbf{r}_1,\,\mathbf{r}_2)=0$. 
 The system's mass (\ref{ClasE2B}) in the c.m. frame is Lorentz-scalar. 
 In case of free particles ($\mathsf{V}=0$) the invariant mass 
$\mathsf{M}=\sqrt{\mathbf{p}^2+m_1^2}+\sqrt{\mathbf{p}^2+m_2^2}$ can be 
transformed for $\mathbf{p}^2$ as 
\be
\mathbf{p}^2
=\frac 1{4s}(s-m_-^2)(s-m_+^2)\equiv\mathsf{k}^2,
\ee{InvMom1}
which is relativistic invariant, $s=\mathsf{M}^2$ is the Mandelstam's 
invariant, $m_-=m_1-m_2$, $m_+=m_1+m_2$. 

 Equation (\ref{TwoEnr}) is the zeroth component of the 4-vector 
(\ref{Main4vec}) and the potential $\mathsf{W_0}$ is Lorentz-vector. 
 But, in the c.m. frame the mass (\ref{ClasE2B}) is Lorentz-scalar; 
and what about the potential $\mathsf{V}$? 
 Is it still Lorentz-vector? 
 To show that the potential is Lorentz-scalar, let us reconsider 
(\ref{ClasE2B}) as follows. 
 The relativistic total energy $\epsilon_i(\mathbf{p})$ ($i=1,\,2$) 
of particles in (\ref{ClasE2B}) given by 
$\epsilon_i^2(\mathbf{p})=\mathbf{p}^2+m_i^2$ can be represented as 
sum of the kinetic energy $\tau_i(\mathbf{p})$ and the particle rest 
mass $m_i$, i.\,e., $\epsilon_i(\mathbf{p})=\tau_i(\mathbf{p})+m_i$. 
 Then the system's total energy (invariant mass) (\ref{ClasE2B}) can be 
written in the form $\mathsf{M}=\sqrt{\mathbf{p}^2+\mathsf{m}_1^2(r)}
+\sqrt{\mathbf{p}^2+\mathsf{m}_2^2(r)}$, where 
$\mathsf{m}_{1,2}(r)=m_{1,2}+\frac 12\mathsf{V}(r)$ are 
the distance-dependent particle masses~\ci{MyNDA17} and (\ref{InvMom1}) 
with the use of $\mathsf{m}_1(r)$ and $\mathsf{m}_2(r)$ takes the form, 
\be
\mathbf{p}^2 = 
K(s)\left[s-(m_+ +\mathsf{V})^2\right]\equiv\mathsf{k}^2-U(s,\,r),
\ee{InvMom2}
where $K(s)=(s-m_-^2)/4s$, $\mathsf{k}^2$ is squared invariant 
momentum given by (\ref{InvMom1}) and 
$U(s,\,r) = K(s)\left[2m_+\mathsf{V} +\mathsf{V}^2\right]$ 
is the potential function. 
 The equation (\ref{InvMom2}) is the relativistic analogy of the NR 
expression $\mathbf{p}^2=2\mu[E-V(r)]\equiv\mathsf{k}^2-U(E,r)$. 

 The equality (\ref{InvMom2}) with the help of the fundamental 
correspondence principle gives the two-particle spinless wave 
equation,
\be 
\left[\left(-i\vec\nabla\right)^2 +\mathsf{k}^2 
-U(s,\,r)\right]\psi(\mathbf{r})=0. 
\ee{Rel2Eq}
 The equation (\ref{Rel2Eq}) can not be solved by known methods for 
the potential (\ref{VmodCor}). 
 Here we use the quasiclassical (QC) method and solve another wave 
equation~\ci{MyMPLA97,MyPRA96}. 
 Compare (\ref{Rel2Eq}) with the one (\ref{TodorEq}). 
 Is there any difference between them?

\section{ Solution of the QC wave equation}\label{SolQCEq}
 Solution of the Shr\"odinger-type's wave equation (\ref{Rel2Eq}) 
can be found by the QC method developed in~\ci{MyPRA96}. 
 In our method one solves the QC wave equation derivation of which 
is reduced to replacement of the operator $\vec{\nabla}^2$ in 
(\ref{Rel2Eq}) by the canonical operator $\Delta^c$ without the first 
derivatives, acting onto the state function 
$\Psi(\vec r)=\sqrt{{\rm det}\,g_{ij}}\psi(\vec r)$, where $g_{ij}$ 
is the metric tensor. 
 Thus, instesd of (\ref{Rel2Eq}) one solves the QC equation, for 
the potential~(\ref{VmodCor}), 
\be 
\Biggl\{\frac{\pa^2}{\pa r^2}+\frac 1{r^2}\frac{\pa^2}{\pa\theta^2}
+\frac 1{r^2\sin^2\theta}\frac{\pa^2}{\pa\varphi^2}
+\frac{s-m_-^2}{4s}\biggl[s-\left(m_+ -\frac 43\frac{\alpha_\S(r)}r
+\sigma r\right)^2\biggr]\Biggr\}\Psi(\mathbf{r})=0.
\ee{Rel2Equa}
 This equation is separated. 
 Solution of the angular equation was obtained in~\ci{MyPRA96} by 
the QC method in the complex plane, that gives 
$\textrm{M}_l=(l+\frac 12)\hbar$, for the angular momentum eigenvalues. 
 These angular eigenmomenta are universal for all spherically symmetric 
potentials in relativistic and NR cases. 

 The radial problem has four turning points and cannot be solved by 
standard methods. 
 We consider the problem separately by the QC method for the short-range 
Coulomb term (heavy mesons) and the long-range linear term (light mesons). 
 The QC method reproduces the exact energy eigenvalues for all known 
solvable problems in quantum mechanics~\ci{MyMPLA97,MyPRA96}. 
 The radial QC wave equation of (\ref{Rel2Equa}) for the Coulomb term 
has two turning points and the phase-space integral is found in 
the complex plane with the use of the residue theory and method of 
stereographic projection~\ci{MyPRA96,MyAHEP13} that gives 
\be
\mathsf{M}_\N^2 
=\left(\sqrt{\epsilon_\N^2}\pm \sqrt{(\epsilon_\N^2)^*}\right)^2 
\equiv 4\left[\Re\{\epsilon_\N^2\}\pm i\Im\{\epsilon_\N^2\} \right],
\ee{W2Coul}
where $\epsilon_\N^2
=\frac 14 m_+^2\left(1-v_\N^2\right) +\frac i2 m_+ m_-v_\N$, 
$v_\N=\frac 23\alpha_\infty/N$, $N=k+l+1$. 

 Large distances in hadron physics are related to the problem of 
confinement. 
 The radial problem of (\ref{Rel2Equa}) for the linear term has four 
turning points, i.\,e., two cuts between these points. 
 The phase-space integral in this case is found by the same method 
of stereographic projection as above that results in the cubic 
equation~\ci{MyNDA17}: $s^3 + a_1s^2 + a_2s + a_3 = 0$, where 
$a_1=16\tilde\alpha_\infty\sigma-m_-^2$, 
$a_2=64\sigma^2\left(\tilde\alpha_\infty^2-\tilde N^2
-\tilde\alpha_\infty m_-^2/4\sigma\right)$, 
$a_3=-(8\tilde\alpha_\infty\sigma m_-)^2$, $\tilde N=N+k+\frac 12$, 
$\tilde\alpha_\infty=\frac 43\alpha_\infty$. 
 The first root $s_1(N)$ of this equation gives the physical solution 
(complex eigenmasses), $\mathsf{M}_1^2(N)=s_1(N)$, for the squared 
invariant mass. 

 Two exact asymptotic solutions obtained such a way are used to derive 
the interpolating mass formula. 
 The~interpolation procedure for these two solutions~\ci{MyZPhC94} 
is used to derive the resonance's mass formula: 
\be
\mathsf{M}_\N^2 
=\left(m_1+m_2\right)^2 \left(1-v_\N^2\right)\pm 2i(m_1^2-m_2^2)v_\N 
+\Re\{\mathsf{M}_1^2(N)\}.
\ee{W2int}
 The real part of the square root of (\ref{W2int}) defines the~centered 
masses and its imaginary part defines the~total widths, 
$\Gamma_\N^\TOT=-2\,\Im\{\mathsf{M}_\N\}$, of 
resonances~\ci{MyAHEP13,MyNPCS14}. 
 The real-part mass in (\ref{W2int}) exactly coincides with the universal mass formula obtained independently by another method with 
the use of the two-point Pad\'e approximant~\ci{MyZPhC94} and is 
very transparent physically, as well as the Coulomb potential. 

 The free fit to the data show a~good agreement for the light and heavy 
$Q\bar q$ meson resonances. 
Note, that the gluon mass in the independent fitting is the same, 
$m_g=416$\,MeV. 
Besides, it is the same for glueballs~\ci{MyEPJC12}. 
the $d$ quark effective mass is also practically the same, i.\,e., 
$m_d\simeq 273$\,MeV, for the light and heavy resonances. 
\begin{table}[ht]
\begin{center}
\caption{The masses of the $\rho^\pm(u\bar d)$-meson resonances}
\label{rho_mes}
\begin{tabular}{lllll}
\hline\noalign{\smallskip}
\ \ Meson &~~~$J^{PC}$ &~~~$\ \ E_n^{ex}$ &~~~$\ \ E_n^{th}$&
~~~Parameters in (\ref{W2int})\\
\noalign{\smallskip}\hline\hline\noalign{\smallskip}
\ \ \ $\rho\ (1S)$&~~~$1^{--}$&~~~$\ \ 776$&~~~$\ \ 776$&~~~~~$\Lambda=500$ MeV\\
\ \ \ $a_2(1P)$&~~~$2^{++}$&~~~$\ 1318$&~~~$\ 1314$&~~~~~$\mu_g=416$\,MeV\\ 
\ \ \ $\rho_3(1D)$&~~~$3^{--}$&~~~$\ 1689$&~~~$\ 1689$&~~~~~$\sigma=0.139$\,GeV$^2$\\ 
\ \ \ $a_4(1F)$&~~~$4^{++}$&~~~$\ 1996$&~~~$\ 1993$&~~~~~$m_d=276$\,MeV\\ 
\ \ \ $\rho\ (1G)$&~~~$5^{--}$&~~~$ ~ $&~~~$\ 2255$&~~~~~$m_u=129$\,MeV\\
\ \ \ $\rho\ (2S)$&~~~$1^{--}$&~~~$\ 1717$&~~~$\ 1682$&~\\
\ \ \ $\rho\ (2P)$&~~~$2^{++}$&~~~$ ~ $&~~~$\  1990$&~\\
\ \ \ $\rho\ (2D)$&~~~$3^{--}$&~~~$ ~ $&~~~$\ 2254$&~\\
\noalign{\smallskip}\hline
\end{tabular}
\caption{The masses of the $D^{*\pm}(c\bar d)$-meson resonances}
\label{Dp_mes}
\begin{tabular}{lllll}
\hline\noalign{\smallskip}
\ \ Meson &~~~$J^{PC}$ &~~~$\ \ E_n^{ex}$ &~~~$\ \ E_n^{th}$&
~~~Parameters in (\ref{W2int})\\
\noalign{\smallskip}\hline\hline\noalign{\smallskip}
\ \ \ $D^*(1S)$&~~~$1^{--}$&~~~$\ 2010$&~~~$\ 2010$&~~~~~$\Lambda=446$\,MeV\\
\ \ \ $D_2^*(1P)$&~~~$2^{++}$&~~~$\ 2460$&~~~$\ 2464$&~~~~~$m_g=416$\,MeV\\ 
\ \ \ $D_3^*(1D)$&~~~$3^{--}$&~~~$\ ~ $&~~~$\ 2845$&~~~~~$\sigma=0.249$\,GeV$^2$\\ 
\ \ \ $D_4^*(1F)$&~~~$4^{++}$&~~~$\ ~ $&~~~$\ 3178$&~~~~~$m_c=1163$\,MeV\\
\ \ \ $D_5^*(1G)$&~~~$5^{--}$&~~~$\ ~ $&~~~$\ 3478$&~~~~~$m_d=271$\,MeV\\
\ \ \ $D^*(2S)$&~~~$1^{--}$&~~~$\ 1820$&~~~$\ 2821$&~\\
\ \ \ $D^*(2P)$&~~~$2^{++}$&~~~$\ 2011$&~~~$\ 3166$&~\\
\ \ \ $D^*(2D)$&~~~$3^{--}$&~~~$\ ~ $&~~~$\ 3471$&~\\
\noalign{\smallskip}\hline
\end{tabular}
\end{center}
\end{table}

 It describes equally well the mass spectra of all $q\bar q$ and $Q\bar Q$ mesons ranging from the $u\bar d$ ($d\bar d$, $u\bar u$, $s\bar s$) states up to the heaviest known $b\bar b$ systems~\ci{MyZPhC94} and 
glueballs~\ci{MyEPJC12,MyEPL10} as well. 
 Besides, it allows one to get the Regge trajectories as analytic functions in the whole region from solution of the cubic equation 
for the angular momentum $J(\mc{M}^2)$~\ci{MyZPhC94}; the Regge 
trajectories including the Pomeron~\ci{MyEPJC12,MyEPL10} are ``saturating'' and appears to be successful in many 
applications~\ci{RossiP03,CLAS_PRL03,CLAS_EPJA05}. 
 
 In our QC method not only the total energy, but also momentum of 
a particle-wave in bound state is the {\em constant of motion}. 
 Solution of the QC wave equation in the whole region is written 
in elementary functions as~\ci{MyMPLA97,MyPRA96,MyClaSol03},
\be
\bar{R}_n(r)=C_n\left\lbrace
\ba{cll}
&\frac 1{\sqrt 2}\exp(|p_n|r -\phi_1), &\,r<r_1,\\
&\cos(|p_n|r -\phi_1 -\frac\pi 4), &\,r_1\le r\le r_2,\\
&\frac{(-1)^n}{\sqrt 2}\exp(-|p_n|r +\phi_2), &\,r>r_2,
\ea\right. \ee{osol}
where $C_n=\sqrt{2|p_n|/[\pi(n+\frac 12)+1]}$ is the normalization
coefficient, $p_n$ is the corresponding eigenmomentum,
$\phi_1=-\pi(n+\frac 12)/2$ and $\phi_2=\pi(n+\frac 12)/2$
are the values of the phase-space integral at the turning points
$x_1$ and $x_2$, respectively. 
 In the classically allowed region [$x_1,x_2$], the solution is
\be
\bar{R}_n(r)=C_n\cos\left(|p_n|r+\frac{\pi }2n\right),
\ee{TiR}
i.e., has the form of a standing wave. 
 This solution is appropriate for two-turning-point problems both 
in non-relativistic and relativistic cases with the corresponding 
eigenmomenta $p_n$. 
 We use this fact in the present work. 

To demonstrate its efficiency we calculate the leading-state masses of 
the $\rho$ and $D^*$ meson resonances (see tables, where masses are in MeV). 

\section*{Conclusion} 
\label{Conclu}
The constituent quark picture could be questioned since potential models 
have serious difficulties because the potential is non-relativistic concept. 
However, in spite of non-relativistic phenomenological nature, 
the potential approach is used with success to describe mesons as~bound 
states of quarks. 

 We have modeled meson resonances to be the quasi-stationary states of two quarks interacting by the QCD-inspired funnel-type potential with 
the coordinate dependent strong coupling, $\alpha_\S(r)$. 
Using the complex analysis, we have derived the meson complex-mass 
formula~(\ref{W2int}), in which the real and imaginary parts are exact 
expressions. 
This approach allows to simultaneously describe in the unified way 
the centered masses and total widths of resonances. 
We have shown here the results only for unflavored and charmed meson 
resonances, however, we have obtained a~good description for strange 
and beauty mesons as well~\ci{MesResX17}. 

\bibliography{BiDaQM}

\end{document}